\documentclass[aps,prb,twocolumn,superscriptaddress,showkeys,reprint]{revtex4-2}

\usepackage{graphicx}% Include figure files
\usepackage{dcolumn}% Align table columns on decimal point
\usepackage{bm}% bold math
\usepackage{xfrac}
\usepackage{graphics,epsfig,graphicx}
\usepackage{ulem}
	\usepackage{siunitx}
	\usepackage{epstopdf}
	\usepackage{verbatim}
	\usepackage{amsmath}
	\usepackage[unicode=true, bookmarks=true, bookmarksnumbered=false, bookmarksopen=false, breaklinks=false, pdfborder={0 0 1}, backref=false, colorlinks=false]{hyperref}
	\usepackage{cleveref}

\usepackage{soul} % for strike out text
\DeclareSIUnit{\nT}{\nano\tesla}
\DeclareSIUnit{\uT}{\micro\tesla}
\DeclareSIUnit{\mT}{\milli\tesla}
\DeclareSIUnit{\T}{\tesla}
\DeclareSIUnit{\mK}{\milli\kelvin}

\begin{document}

\title{Threshold and frequency properties of a cold ytterbium laser}

\author{Dmitriy Sholokhov}
\email{dmitriy.sholokhov@physik.uni-saarland.de}
\affiliation{ 
Experimentalphysik, Universität des Saarlandes, D-66123 Saarbrücken, Germany
}%

\author{Saran Shaju}%
\affiliation{ 
Experimentalphysik, Universität des Saarlandes, D-66123 Saarbrücken, Germany
}%
\author{Ke Li}
\affiliation{ 
Experimentalphysik, Universität des Saarlandes, D-66123 Saarbrücken, Germany
}%
\author{Simon B. Jäger}
\affiliation{ 
Physics Department and Research Center OPTIMAS,
University of Kaiserslautern-Landau, D-67663 Kaiserslautern, Germany
}%
\author{Jürgen Eschner}
\email{juergen.eschner@physik.uni-saarland.de}
\affiliation{ 
Experimentalphysik, Universität des Saarlandes, D-66123 Saarbrücken, Germany
}%

\date{\today}

\begin{abstract}
%abstract
We investigate properties of the lasing action observed on the $^1$S$_0\,\leftrightarrow\,^3$P$_1$ intercombination transition of ytterbium atoms that are laser-cooled and -trapped inside a high-finesse cavity. The dressing of the atomic states on the $^1$S$_0\,\leftrightarrow\,^1$P$_1$ transition by the magneto-optical trap (MOT) laser light allows the coupled atom-cavity system to lase, via a two-photon transition, on the same line on which it is pumped. The observation and basic description of this phenomenon was presented earlier by Gothe et al.~\cite{Gothe2019}. In the current work, we focus on a detailed analysis of the lasing threshold and frequency properties and perform a comparison to our theoretical models.
\end{abstract}


\maketitle

%\tableofcontents
%\begin{quotation}
%Mostly place holder texts and images
%\end{quotation}

\label{sec:intro}
\section{Introduction}
The conventional laser, based on population inversion and cavity-stimulated emission between two atomic eigenstates, has been extended to more complex lasing mechanisms such as the cold atom \cite{Mollow1972, Grison_1991, Tabosa1991, Hilico_1992, Guerin2008, Salzburger:2004, Salzburger:2006, Froufe-Perez2009, Guerin2010, Gerasimov2014, Harmon2022}, the inversionless \cite{Javan1957, Lindberg1988, Harris1989, Kitching1999, Mompart_2000} and the super-radiant \cite{Meiser2010, Bohnet2012, Bohnet2014, Norcia2016, Zhang:2018, Bychek2023, Fasser2023} laser. These approaches aim to discover, study, and apply new light-matter interactions \cite{Xu:2016, Norcia2016:2, Jaeger:2017, Norcia:2018, Gothe2019:2, Hotter:2019, Cline2022, Mivehvar2024}, for example to better understand collective dynamics and matter phases in cold atom gases. New advances in this direction will also help to overcome thermal-noise-limited laser linewidth \cite{Ludlow2007, Dawel2023, Bychek2023}, providing new tools for the atomic clock world, such as gravimeters \cite{Yudin2019}, dark matter analysers \cite{Sherrill2023}, or fundamental constants testers \cite{Barontini2022}.

For such purposes, it is particularly promising to realize ultra-narrow tunable lasers, which are based on optical transitions in cold atomic systems~\cite{Meiser2009, Meiser2010, Norcia2016}. Atoms with a narrow transition are thus placed inside of a rapidly decaying cavity, in the so-called "bad-cavity" regime~\cite{Chen2009, Auffeves2010, Zhang2023}. In this regime the laser linewidth is less affected by noise in the cavity degrees of freedom since the coherence is mostly stored in the atomic degrees of freedom. Using the forbidden transitions of cooled and trapped atoms would allow one to achieve lasing linewidth on the \SI{1}{\milli\hertz} scale~\cite{Chen2009, Meiser2009, Meiser2010, Hotter2020}. 

Lasing mechanisms in cold atomic systems have been studied with rubidium~\cite{Guerin2008, Bohnet2012}, calcium~\cite{Wang:2012, Laske:2019}, 
%cesium, 
strontium~\cite{Meiser2009, Meiser2010, Norcia2016, Norcia:2018:2, Debnath:2018}, and ytterbium~\cite{Gothe2019}. Recent progress towards a super-radiant laser has been obtained with strontium atoms~\cite{Schaeffer2020, Liu:2020, Jaeger:2021, Zhang:2021, Tang2022, Cline2022, Kristensen2023, Niu23}. Cold ytterbium (Yb) is an equally promising platform, in particular with respect to its ultra-narrow clock transition $^1\text{S}_0\,\leftrightarrow\,^3\text{P}_0$ ($\sim\SI{10}{\milli\hertz}$ linewidth). In related work, we recently demonstrated a cold-atom laser in continuous-wave operation on the semi-forbidden $^1\text{S}_0\,\leftrightarrow\,^3\text{P}_1$ transition ($\sim\SI{182}{\kilo\hertz}$ linewidth)~\cite{Gothe2019}.

In that work we identified a laser gain mechanism based on a two-photon process that involves the light used for trapping and cooling the atoms in a magneto-optical trap (MOT). The two-photon transition effectively couples the $^3$P$_1$ and $^1$P$_1$ atomic states through simultaneous photon emission into the cavity and photon absorption from the MOT light. Population inversion is established because of the large difference in the $^1$P$_1$ and $^3$P$_1$ lifetimes. The cavity with a decay time on the order of the $^3$P$_1$ lifetime keeps the system in between the good- and bad-cavity regimes and supports the described process. 

While we established the basic mechanism, the detailed parameter regimes required for the lasing have been far less understood. This was partially due to the absence of a qualitative and quantitative theory that can predict the lasing threshold and behavior. In this paper, we close these gaps and provide a detailed analysis of the lasing threshold and frequency, both as functions of the parameters of the exciting (pump) light, the MOT light, and the cavity. With two different approaches based on optical Bloch equations~\cite{Harmon2022}, we model steady-state and dynamical properties of the laser, and we compare the results to our experimental findings, finding excellent qualitative and good quantitative agreement. 

In the next Section II, we introduce our experimental setup. We continue by describing the performed measurements and their results in Sec.~III. In Sec.~IV and V we present an analysis of threshold and frequency properties of the cold ytterbium laser and compare them with our models. We conclude and give an outlook in Sec.~VI.

\section{Experiment}
\label{sec:experiment}

\begin{figure*}[!ht]
	\centering
	\includegraphics[width=0.95\textwidth]{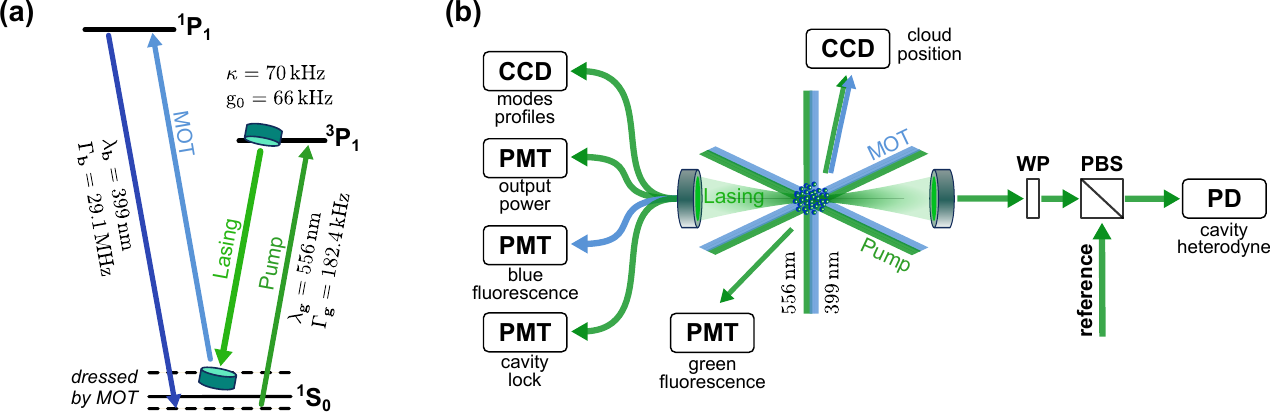}
	\caption{(a) Energy levels and transitions of atomic $^{174}\mathrm{Yb}$ with relevant parameters. Cavity parameters are also indicated. (b) Schematic of magneto optical trap geometry, cavity interaction, and detection pathways.}
	\label{fig:1_setup}
\end{figure*}

Our laser is based on a magneto-optically trapped cloud of $^{174}$Yb atoms inside a high-finesse cavity. Atoms are evaporated from an oven operating at $\qty{500}{\degreeCelsius}$ and collimated. While travelling towards the trapping region, they are decelerated from $\sim\qty{300}{\meter\per\second}$ to $\sim\qty{1}{\meter\per\second}$ by a Zeeman slower. The MOT operates on the $\Gamma_\textbf{b} = \SI{29.1}{\mega\hertz}$-wide $^1$S$_0\,\leftrightarrow\,^1$P$_1$ transition at $\lambda_\textbf{b} =\SI{399}{\nano\meter}$ wavelength (blue), with a typical detuning of $\Delta_\text{MOT} \simeq -\qty{30}{MHz}$ and with $38\,\mathrm{G/cm}$ axial magnetic field gradient. [Note that throughout the paper, frequencies are given in technical units. The corresponding angular frequencies are used in the theoretical analyses.] It traps and cools the atoms in a $\sim\qty{1}{mm}$-size spatial region around the center of our optical cavity, see Fig.~\ref{fig:1_setup}. The number of trapped atoms is controlled by a servo-driven mechanical shutter that regulates the atomic-beam flux. We trap up to $10^7$ atoms in the MOT at 5 to $\qty{10}{mK}$ temperature. 

The $\SI{4.78}{\centi\meter}$-long cavity with $\SI{87}{\micro\meter}$ waist radius is resonant with the $\Gamma_\textbf{g} = \qty{182.4}{\kilo\hertz}$-wide $^1$S$_0\,\leftrightarrow\,^3$P$_1$ intercombination transition at $\lambda_\textbf{g} = \SI{556}{\nano\meter}$ wavelength (green). It has an energy decay rate $\kappa = \SI{70}{\kilo\hertz}$ (finesse $\mathcal{F}=$~\qty{45000}) and a vacuum Rabi frequency $g_0 = \SI{66}{\kilo\hertz}$ (cooperativity per atom $C_1\simeq0.34$). Taking into account the ratio between the cloud size and the cavity waist, we estimate that $N = 10^4$ to $10^5$ of atoms interact with the TEM$_{00}$ cavity mode, placing the system well within the collective strong-coupling regime $C = NC_1 \gg 1$~\cite{Savant2017}. The cavity is rotated by $45^\circ$ with respect to the in-plane MOT beams, as schematically depicted in Fig.~\ref{fig:1_setup}(b). It is frequency-locked by means of the TEM$_{15,0}$ cavity mode detuned by \SI{-420}{\mega\hertz} from TEM$_{00}$.

The trapped atomic cloud is additionally laser-excited (pumped) on the green $^1$S$_0\,\leftrightarrow\,^3$P$_1$ line by six beams spatially overlapped with the blue MOT beams, as depicted in Fig.~\ref{fig:1_setup}(b). For appropriate experimental parameters, the light that the atoms emit into the cavity builds up to a large intra-cavity field. About $T_\text{m} \simeq \qty{1.5}{ppm}$ per mirror of the intra-cavity power leaks out and is observed as laser output~\cite{Gothe2019}. 

We characterize the output with different techniques, see Fig.~\ref{fig:1_setup}(b): most importantly, a heterodyne measurement is performed by mixing the cavity TEM$_{00}$ mode output with a \qty{30}{MHz}-detuned reference beam. The beat note is detected by a fast photoreceiver (New Focus) and measured on a spectrum analyzer (Rohde $\&$ Schwarz). The local oscillator of the heterodyne measurement and the excitation light are derived from the same green laser using different acousto-optical modulators (AOMs), hence the spectrum analyzer reveals any frequency shift of the laser output relative to the pump. Additionally, the structure of the cavity mode is detected by a CCD-camera (Imaging Source), and the fluorescence on the green and blue transitions and the power emitted from the cavity are measured by photomultiplier tubes (PMTs, Hamamatsu); these measurements are presented in the supplementary material. 

It should be noted that the six-beam configuration of the green pump light creates an additional trapping and cooling effect ("green MOT"), co-acting or counter-acting to the blue MOT, depending on its detuning relative to the green resonance. When co-acting at negative detuning, the green MOT leads to lower cloud temperatures, and hence higher atomic densities. On the other hand, at positive detuning, the green light is heating the cloud, thus reducing the trapping and cooling action of the blue MOT and depleting the number of atoms interacting with the cavity mode. In a comprehensive model, this mechanical effect would have to be taken into account; here we use a simplified treatment that neglects this contribution. 

\section{Measurement}

The raw data that we record to characterize the lasing properties are obtained from the heterodyne measurement of the TEM$_{00}$ mode emission for a wide range of cavity detunings $\Delta_\text{cavity}$, green pump and blue MOT detunings, $\Delta_\text{pump}$ and $\Delta_\text{MOT}$, and pump and MOT powers, $P_\text{pump}$ and $P_\text{MOT}$. Detunings are always referenced to the respective atomic line centers. 

For a given set of parameters, we record the spectrum of the beat signal from the photodiode on one side of the cavity with the rf spectrum analyser. Simultaneously, we record the cavity output power and the mode structure observed with a CCD camera on the opposite side of the cavity, see Fig.~\ref{fig:2_powers}(b). The atomic fluorescence signals on the blue and green transitions are measured with an oscilloscope.
For a given value of $\Delta_\text{cavity}$, we step $\Delta_\text{pump}$, while $\Delta_\text{MOT}$ as well as $P_\text{pump}$ and $P_\text{MOT}$ are kept constant. The measured heterodyne beat spectra are transformed into two types of heatmaps (color maps): one representing the cavity output power, as obtained from the peak area of the heterodyne signal, the other representing the output frequency, as obtained from the peak position. Both maps are then rendered, with pump detuning as X-axis and cavity detuning as Y-axis. These maps are generated for each set of values of $\Delta_\text{MOT}$, $P_\text{pump}$ and $P_\text{MOT}$. Details of the data processing will be provided as supplementary information. The camera data are saved as video recordings for additional analyses.

\section{Threshold}

\begin{figure}[!ht]
	\centering
	\includegraphics[width=0.45\textwidth]{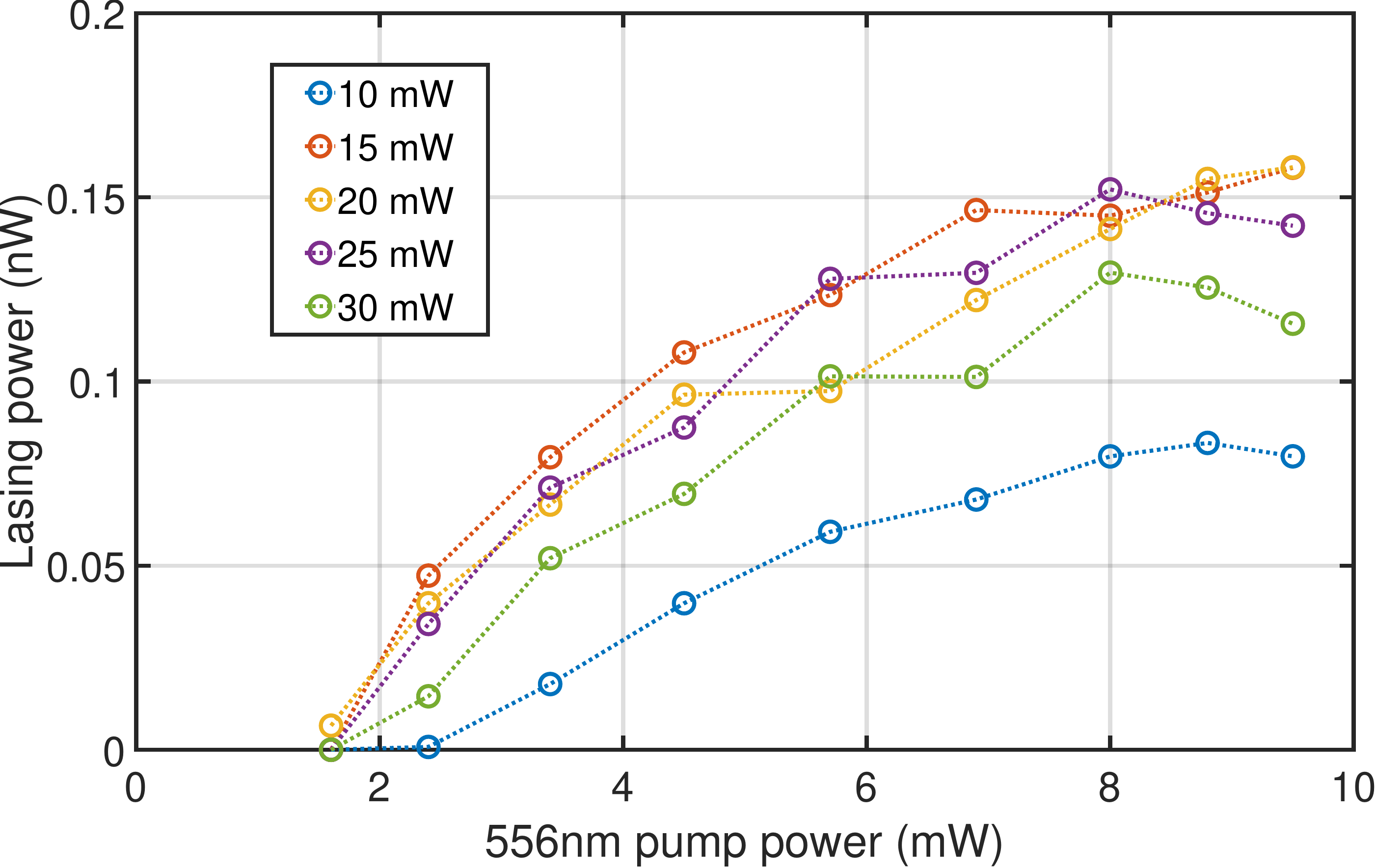}
	\caption{Output power of the cold Yb laser as a function of the green pump power for several values of blue MOT power. }
	\label{fig:2_powers}
\end{figure}

\begin{figure*}[!ht]
	\centering
	\includegraphics[width=1\textwidth]{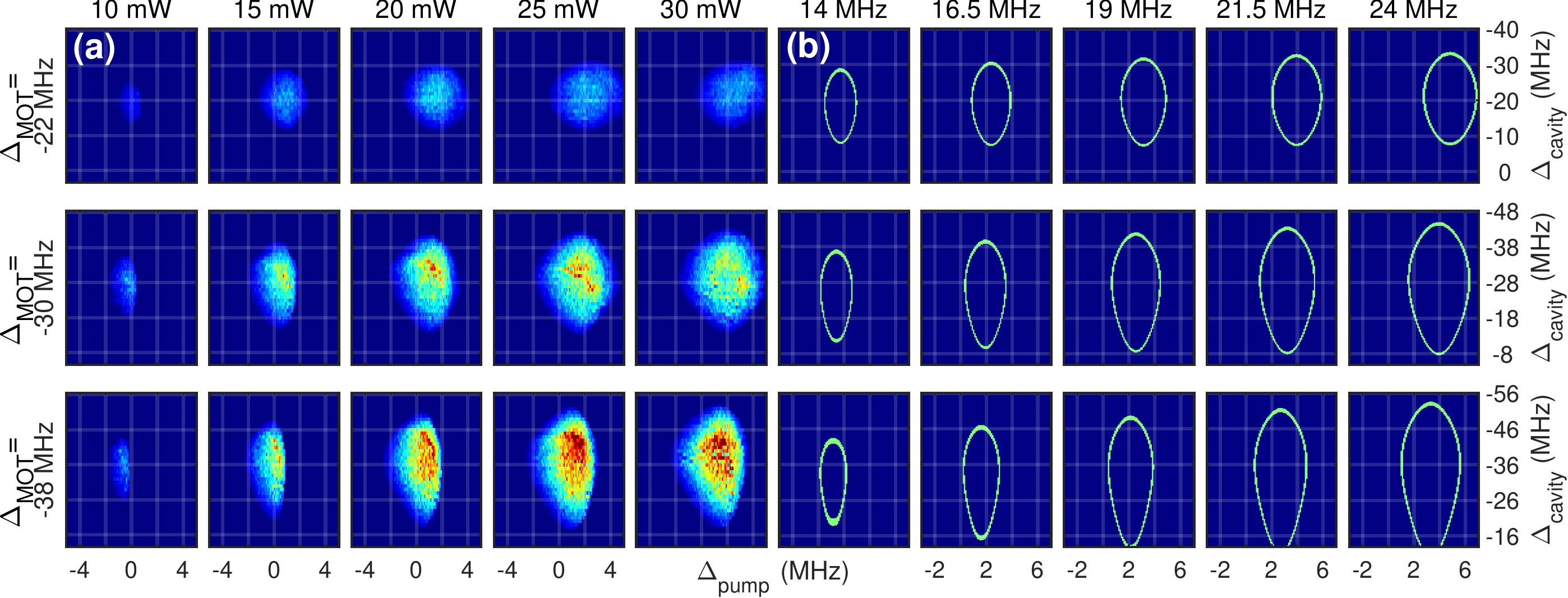}
	\caption{Laser output power and threshold. (a) Measured laser power, color coded from dark blue (zero) to dark red (maximum). (b) Simulated laser threshold shown as contours; outside the contours, the model predicts no lasing, inside it predicts laser output. In both (a) and (b), each sub-plot is measured resp.\ calculated as a function of $\Delta_\text{pump}$ (along X) and $\Delta_\text{cavity}$ (along Y); the other variables are $\Delta_\text{MOT}$ (rows) and $P_\text{MOT}$ (columns). MOT power values are given in \si{\milli\watt} for experimental data in (a); for the simulation in (b), the Rabi frequencies in \si{\mega\hertz} are displayed, which are in good agreement with experimentally calibrated ones.}
	\label{fig:3_int_maps}
\end{figure*}

As a first step we illustrate the typical lasing threshold behaviour of our system.  In Fig.~\ref{fig:2_powers}, we display the continuous-wave laser output power as a function of pump power $P_\text{pump}$, for various values of $P_\text{MOT}$ as indicated in the inset. The detunings $\Delta_\text{pump} = 0$, $\Delta_\text{MOT} = -\SI{30}{\mega\hertz}$, and $\Delta_\text{cavity} = -\SI{30}{\mega\hertz}$ are fixed.
We observe typical laser characteristics with a threshold and with output power that increases approximately linearly above threshold. The threshold is in the range of $P_\text{pump} \simeq \qty{2}{mW}$ for all displayed situations. A systematic dependence of the threshold on the detunings is discussed in the following in Fig.~\ref{fig:3_int_maps}. There, the pump power is fixed to $P_\text{pump} = \qty{5.7}{mW}$, for which all curves in Fig.~\ref{fig:2_powers} are well above threshold. 

We now explore the lasing behavior as function of the various  detunings. Figure~\ref{fig:3_int_maps}(a) shows the measured laser output power, as color scale maps, vs.\ the detunings $\Delta_\text{pump}$ and $\Delta_\text{cavity}$. The MOT power and detuning values are indicated at the sub-plots. We observe distinct regions of lasing centered around $\Delta_\text{pump} \sim \SI{0}{\mega\hertz}$ and $\Delta_\text{cavity} \sim \Delta_\text{MOT}$. 
In our earlier work \cite{Gothe2019}, we observed up to three regions of lasing emission, corresponding to $\sigma^\pm$ and $\pi$ transitions, that were separated due to an offset magnetic field in the MOT. Here we see only one lasing region, as a consequence of better alignment of the atomic cloud to the center of the MOT, where the magnetic field is zero. 

In the maps of Fig.~\ref{fig:3_int_maps}(a) we observe a systematic shift of the lasing regions along the $\Delta_\text{pump}$ axis, when $P_\text{MOT}$ is varied, and along the $\Delta_\text{cavity}$ axis, when $\Delta_\text{MOT}$ changes. That dependence is a consequence of the MOT light field dressing the $^1\text{S}_0$ atomic ground state: the state splits into the lower dressed state, from which the atoms are pumped to $^3\text{P}_1$, and the upper dressed state, which is the intermediate state in the two-photon lasing process, as depicted in Fig.~\ref{fig:1_setup}(a). Hence the observed systematics confirm and elaborate the gain mechanism proposed in \cite{Gothe2019}: an atom is first pumped from the lower dressed $^1\text{S}_0$ state to $^3\text{P}_1$; it emits a green photon into the cavity mode and absorbs a blue MOT photon, transiting through the upper dressed state; spontaneous decay from $^1\text{P}_1$ back to the ground state closes the cycle. 

We perform a simulation of our system to describe our experimental results in terms of the lasing threshold. To that end, we calculate the steady state of a three-level Bloch equation model that treats the pump as incoherent drive, while the atom-cavity interaction is treated as coherent, with a collective Rabi frequency $\Omega_\textbf{cavity} = g_0\sqrt{N}$. The simulation shown in Fig.~\ref{fig:3_int_maps}(b) was performed for an atom number $N={75000}$. The other parameters are varied according to the measurement shown in Fig.~\ref{fig:3_int_maps}(a); the green Rabi frequency is fixed to $\Omega_\text{pump} = \SI{1.5}{\mega\hertz}$ which corresponds, according to our calibration, to $P_\text{pump} \simeq \qty{5.7}{mW}$. 
We extract the threshold values from the simulation by equating the calculated gain rate with the cavity loss rate, and we find the contours displayed in Fig.~\ref{fig:3_int_maps}(b). The calculated contours agree well with the observed boundaries of the lasing regions shown in Fig.~\ref{fig:3_int_maps}(a). Inside the contours the model predicts lasing, but in view of the strong simplifications, we do not expect it to provide a reliable value for the lasing power, hence we do not attempt a comparison to the measurement in that regime.

Apart from the threshold contours, two important features of the experimental results are reproduced by the model: (i) the shift of the lasing regions towards larger pump frequencies, when the MOT power increases, which we attribute to the MOT light-induced Stark shift of the ground state; and (ii) the shift of the regions when the MOT frequency is varied, such that the center is always near $\Delta_\text{cavity} \simeq \Delta_\text{MOT}$, which is a signature of the two-photon gain mechanism.  

At the same time, one sees a discrepancy between the ranges of pump detuning for which lasing occurs in the experiment and in the simulation, visible as a systematic difference of $\sim+\SI{2}{\mega\hertz}$ (see the horizontal axes of Fig.~\ref{fig:3_int_maps}(a) and (b)). This is partially attributed to an inaccuracy in the laser lock calibration, partially its origin is not yet understood. 
An obvious source of discrepancies between model and experiment is that the number of atoms interacting with the cavity is assumed constant in the simulations, whereas it varies in the measurement, even within one sub-plot of Fig.~\ref{fig:3_int_maps}(a). The reason is the already mentioned frequency-dependent impact of the pump light on the trapping (the green MOT), which varies from cooling and compressing the atomic cloud to heating and diluting it, as the pump light is tuned across the (Stark-shifted) resonance. One indication for this effect is the sudden transition from lasing to no-lasing on the positive pump frequency side of the lasing regions. 

\section{Frequency}

\begin{figure}[!ht]
	\centering
	\includegraphics[width=0.46\textwidth]{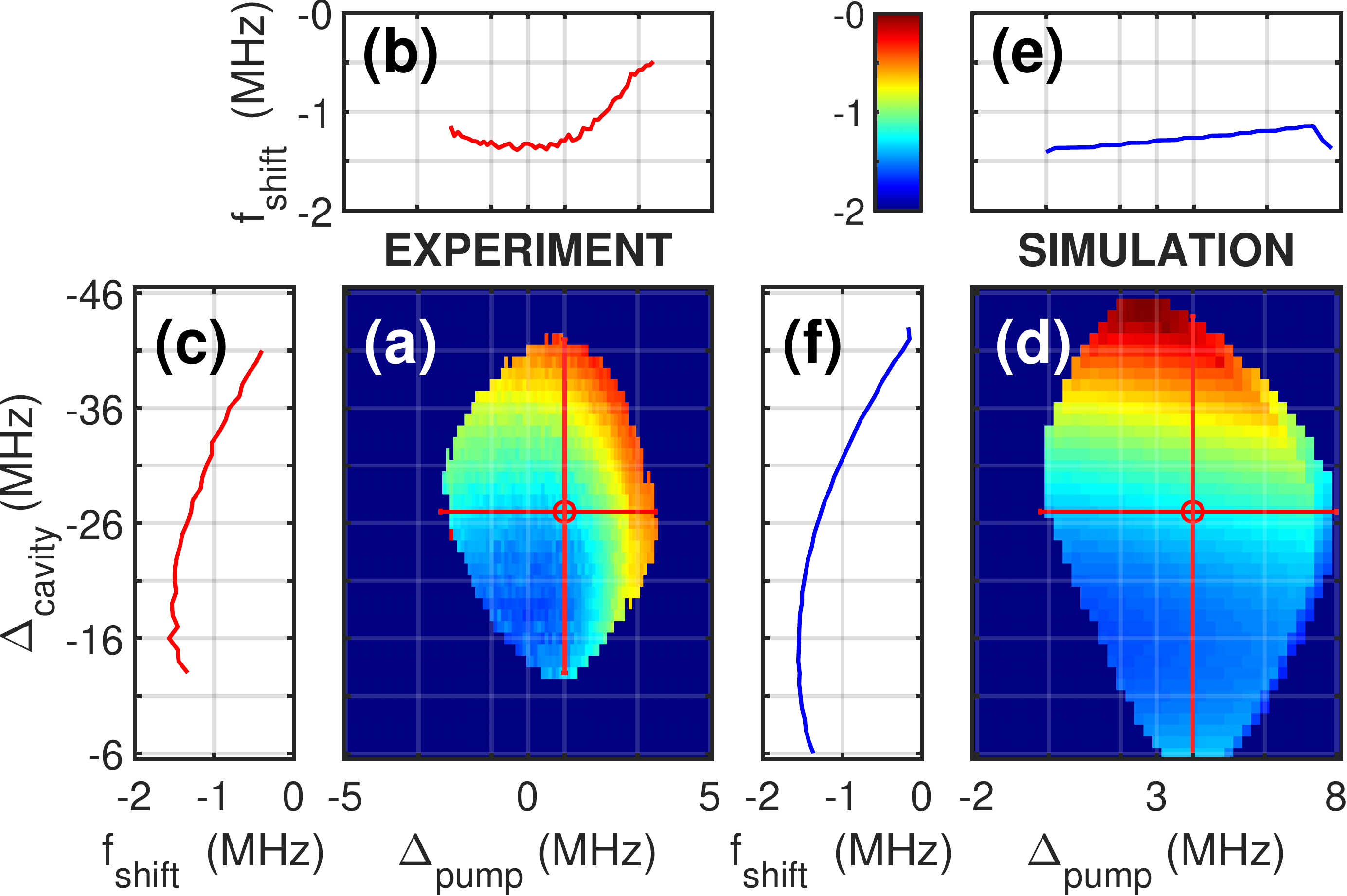}
	\caption{Frequency shift map of the cavity output, measured (a) and simulated (d). The vertical and horizontal cuts, taken along the red lines, depict examples for the frequency shift dependencies on $\Delta_\text{cavity}$, (c) and (f), and on $\Delta_\text{pump}$, (b) and (e). See text for experimental and model parameters.}
	\label{fig:4_light_shift_freq}
\end{figure}

When analysing the frequency of the laser light, obtained via the heterodyne measurement, we observe a systematic redshift with respect to the empty-cavity frequency, the latter set by the cavity lock. We associate this shift with the change of the refractive index inside the cavity due to the presence of the atoms. The amount of redshift depends on the pump and cavity detunings and varies in a rather complex manner over the lasing regions (Fig.~\ref{fig:4_light_shift_freq}(a)). While an intuitive understanding is hard to attain, we carried out a simulation based on mean-field theory~\cite{Harmon2022} and find that it reproduces a similar detuning dependence. The model dynamically calculates the cavity field, and a Fourier transform provides its frequency relative to the empty cavity. 

A comparison of measured and simulated frequency shifts is displayed in the maps of Fig.~\ref{fig:4_light_shift_freq}, where (a) shows the experimental data for $\Delta_\text{MOT} = \SI{-30}{\mega\hertz}$, $P_\text{MOT} = \SI{20}{\milli\watt}$, and $P_\text{pump} = \SI{5.7}{\milli\watt}$, and (d) shows the model result for the corresponding parameters and an atom number $N={75000}$. The Rabi frequencies of the MOT and pump light are $\Omega_\text{MOT} = \SI{19}{\mega\hertz}$ and $\Omega_\text{pump} = \SI{1.5}{\mega\hertz}$, respectively, in good agreement with an independent calibration of the experimental values. Despite the simplifications of the model, it reproduces the main features of the experimental results to a good degree, in particular the absolute value of the frequency shift, up to \SI{-1.5}{\mega\hertz} and its dependence on $\Delta_\text{cavity}$, shown by the vertical cuts in Figs.~\ref{fig:4_light_shift_freq}(c) and (f). 

We note that this model does not take into account atomic motion nor the cooling and trapping action of the green and blue light, as possible contributions to the frequency shift. Corresponding extension of the theory will allow us to investigate the finer details and obtain a deeper understanding of the phenomena.

\section{Conclusion}

We have systematically studied the laser emission from cold $^{174}$Yb atoms  in a high-finesse cavity, on the semi-forbidden $^3$P$_1$ to $^1$S$_0$ transition. Using heterodyne measurement of the laser output, we record the power and the frequency of the emission as functions of the detunings and powers of the involved light fields. We map the parameter regions where lasing occurs and study their systematics, in particular their shifts with the frequency and power of the trapping (MOT) light. The behaviour confirms and elaborates our proposed mechanism for the laser gain \cite{Gothe2019}, a two-photon process that couples the $^3$P$_1$ state to the $^1$P$_1$ state. The intermediate level of the process emerges from the $^1$S$_0$ ground state by its dressing with the MOT light. 

For a comparison to model calculations, we have focused on the laser threshold and the laser output frequency, calculated from a Bloch equation model and a mean-field dynamical model, respectively. Despite significant simplifications, the models reproduce main observed features well, thus corroborating our understanding of the system. Some particular discrepancies are also obvious, which will be subject to further investigation.

\section{Acknowledgements}

We gratefully acknowledge funding by the Deutsche Forschungsgemeinschaft (DFG, German Research Foundation) through the Collaborative Research Center TRR-306 "QuCoLiMa", sub-project B04. S.B.J. acknowledges support from the Deutsche Forschungsgemeinschaft (DFG, German Research Foundation) through projects A4 and A5 in TRR-185 “OSCAR”. The authors acknowledge helpful discussions with J. Cooper, F. Fama, G. Harmon,  M. Holland, N. Kukharchyk, G. Morigi, J. Reilly, and S. Schäffer.

%\nocite{*}
\bibliography{bibliography.bib}% Produces the bibliography via BibTeX.

%\newpage
%\newpage
%\newpage
%\onecolumngrid

%\begin{center}

%\includegraphics[page=1,width=1.0\textwidth,height = 1.0\textheight]{Lasing_001_supp.pdf}
%\includegraphics[page=2,width=1.0\textwidth,height = 1.0\textheight]{Lasing_001_supp.pdf}
%\includegraphics[page=3,width=1.0\textwidth,height = 1.0\textheight]{Lasing_001_supp.pdf}
%\includegraphics[page=4,width=1.0\textwidth,height = 1.0\textheight]{Lasing_001_supp.pdf}
%\includegraphics[page=5,width=1.0\textwidth,height = 1.0\textheight]{Lasing_001_supp.pdf}

%\end{center}

%\centering
%\setboolean{@twoside}{false}

%\includepdf[pages=-]{Lasing_001_supp.pdf}

\end{document}